\documentclass[preprint,aps,showpacs,nofootinbib,preprintnumbers,amsmath,amssymb]{revtex4-1}
\usepackage{}
\usepackage{epsfig}
\usepackage{subfigure}
\usepackage{dcolumn}
\usepackage{bm}
\usepackage[usenames ,dvipsnames]{xcolor}
\usepackage{slashed}
\usepackage{graphicx,color}
\usepackage{centernot}

\begin{document}
\title{Search for $XYZ$ states in $\Lambda_b$ decays at the LHCb
}
\author{Y.K. Hsiao and C.Q. Geng}
\affiliation{
Chongqing University of Posts \& Telecommunications, Chongqing, 400065, China\\
Physics Division, National Center for Theoretical Sciences, Hsinchu, Taiwan 300\\
Department of Physics, National Tsing Hua University, Hsinchu, Taiwan 300
}
\date{\today}

\begin{abstract}
We consider $X(3872)$ and $Y(4140)$ as the vector tetraquark states of 
$X_c^0\equiv c\bar c u\bar u(d\bar d)$ and $c\bar c s\bar s$, respectively. 
By connecting $\Lambda_b\to X_c^0\Lambda$ to $B^-\to X_c^0 K^-$,
we predict that the branching ratios of 
$\Lambda_b\to \Lambda(X(3872)^0\to) J/\psi \pi^+\pi^-$ and $\Lambda_b\to \Lambda(Y(4140)\to) J/\psi \phi$
are $(5.2\pm 1.8)\times 10^{-6}$ and $(4.7\pm 2.6)\times 10^{-6}$, which are accessible to the experiments
at the LHCb, respectively.
The measurements of these $\Lambda_b$ modes would be the first experimental evidences for the $XYZ$ states in baryonic decays.
\end{abstract}

\maketitle
\section{introduction}
With the quantum numbers of $J^{PC}=1^{++}$ determined 
by the $B^-\to X(3872)^0 K^-$ decay~\cite{QN_LHCb},
the state of $X(3872)^0$ has been established as one of the $XYZ$ states~\cite{XYZ_Rev},
which are regarded to be exotic due to the non-pure $c\bar c$ components. 
However, it is still a puzzle  whether $X(3872)^0$ is 
really a tetraquark state (four-quark bound state) with the quark content $c\bar c u\bar u(d\bar d)$~\cite{Maiani:2004vq}.
Note that, while there is no sign of 
its charged partner to be the  $c\bar c u\bar d(d \bar u)$ state,
$Y(4140)$ can be a tetraquark consisting of  $c\bar c s\bar s$~\cite{Stancu:2009ka},
of which the quantum numbers of $J^{PC}$ are not experimentally assigned.
As more investigations are apparently needed, 
the study of $X(3872)$ has been restricted in the $B$ decays of $B\to X(3872)^0 K^{(*)}$ 
and $B\to X(3872)^0 K\pi$ with $K\pi$ partly from $K^*$~\cite{pdg,Bala:2015wep},
where the resonant $X^{0}(3872)$ decay channels can be 
$X(3872)^0\to J/\psi \pi^+\pi^-,\,J/\psi \omega$, $J/\psi\gamma$ and $D\bar D^*$.
At present, no other observation has been found beyond the $B$ decays.

On the other hand, being identified as the exotic meson, which could be
the tetraquark~\cite{Maiani:2004vq},
 $D\bar D^*$ molecule~\cite{molecule_A}, 
or  hybrid $c\bar c g$ bound state~\cite{Li:2004sta}, 
the $X(3872)$ state causes the difficulty of the theoretical calculations. 
In this study, we will concentrate on the tetraquark scenario by denoting $X_c^0$ 
to be composed of $c\bar c q\bar q$, where $q\bar q$ can be $u\bar u$, $d\bar d$, or $s\bar s$.
In particular, we take $X(3872)^0$ and $Y(4140)$ as two of these exotic $X_c^0$ states.
Through the $b\to c\bar c s$ transition at the quark level in Fig.~\ref{fig1},
the decays of $B\to (X(3872)^0,\, J/\psi) K$  correspond to
the  processes of the $B\to K$ transition 
with the recoiled charmed mesons of $X(3872)^0$ and $J/\psi$, respectively.
Although the $J/\psi$ formation from the $c\bar c$ currents can be calculated within the framework of QCD,
 the $X(3872)$ one cannot be done at the moment.

However,  it is interesting to see in Fig.~\ref{fig1}
that  all decays of $(B,\,\Lambda_b)\to (X_c^0,\,J/\psi) K$
are originated from the $b\to c\bar c s$ transition at the quark level, and therefore connected.
As a result,
despite the unknown matrix elements of the $X_c^0$ hadronization through the $c\bar c$ currents,
we can relate these decays.
In particular, we can predict the branching ratios of 
$\Lambda_b\to X_c^0 \Lambda$. 
 The experimental searches of these   $\Lambda_b$  decays at the LHCb
will clearly improve our understanding of the XYZ states.

\section{formalism}
\begin{figure}[t!]
\centering
\includegraphics[width=2.2in]{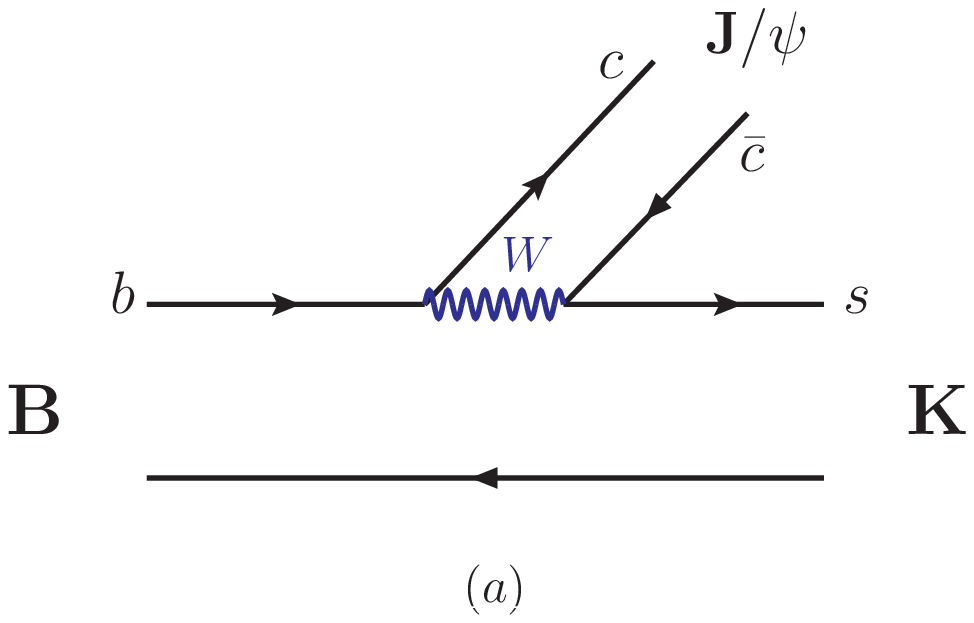}
\includegraphics[width=2.2in]{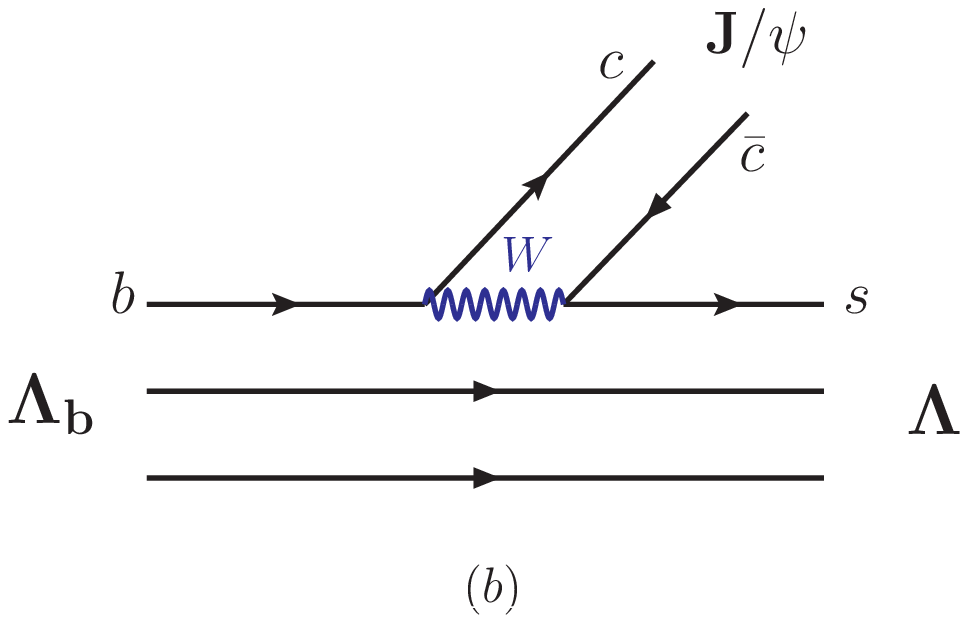}
\includegraphics[width=2.2in]{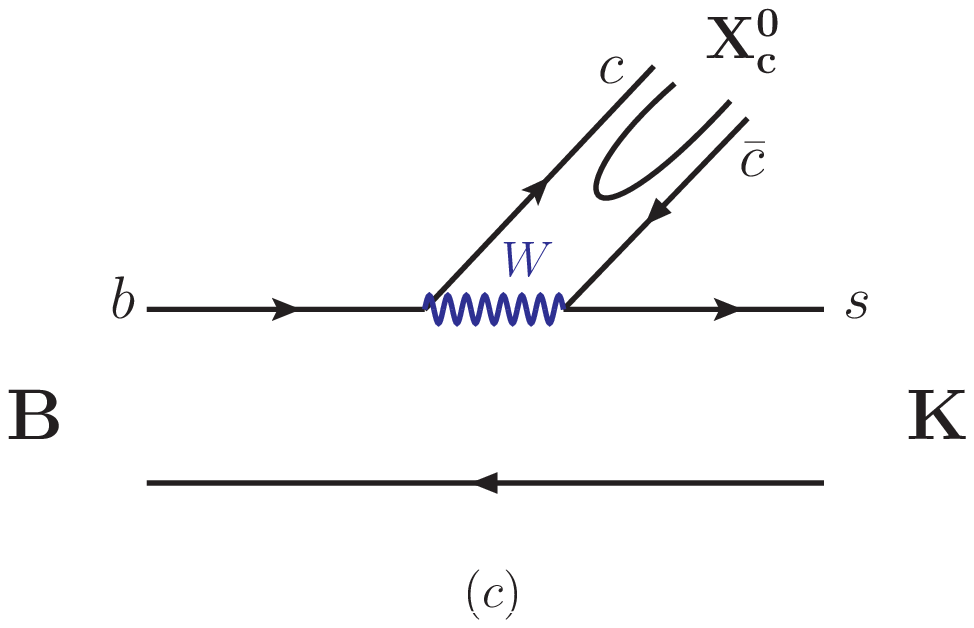}
\includegraphics[width=2.2in]{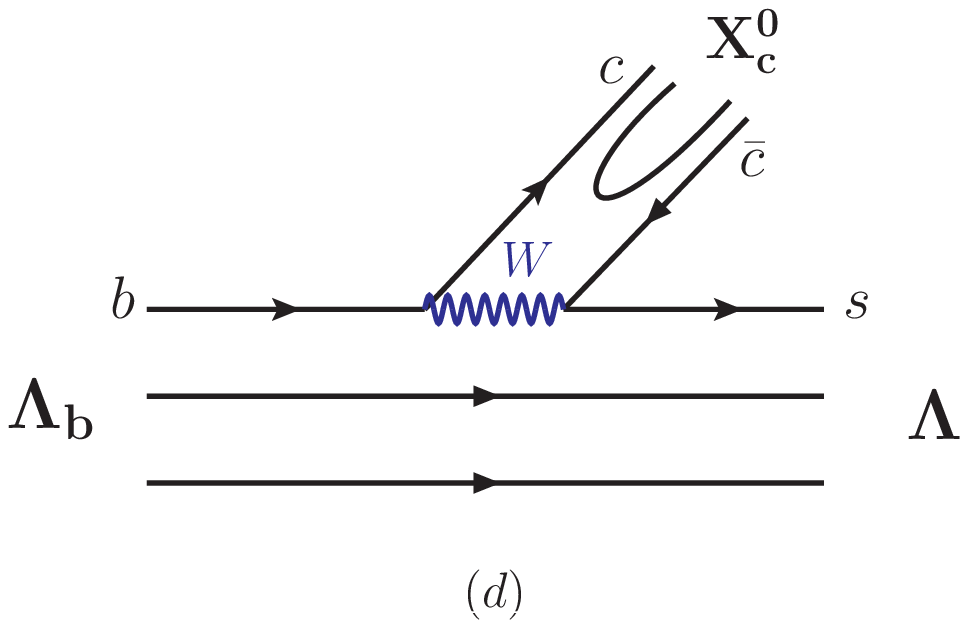}
\caption{
The doubly charmful $b$-hadron decays, where (a), (b), (c), and (d) depict
$B\to J/\psi K$, $\Lambda_b\to J/\psi\Lambda$,
$B\to X_c^0K$, and $\Lambda_b\to X_c^0\Lambda$, respectively, with $X_c^0$ 
as the tetraquark  to consist of $c\bar c q\bar q$.}\label{fig1}
\end{figure}

 From Fig.~\ref{fig1}, 
through the effective Hamiltonian of the $b\to c\bar c s$ transition at the quark level,
the amplitudes of  $\Lambda_b\to M_c \Lambda$ and $B\to M_c K$
can be factorized as~\cite{Hsiao:2015cda,FA_Xc} 
\begin{eqnarray}\label{eq1}
{\cal A}(\Lambda_b\to M_c \Lambda)&=&
\frac{G_F}{\sqrt 2}V_{cb}V_{cs}^*a_2\,
\langle M_c|\bar c\gamma^\mu(1- \gamma_5) c|0\rangle
\langle \Lambda|\bar s\gamma_\mu(1-\gamma_5) b|\Lambda_b\rangle\,,\nonumber\\
{\cal A}(B\to M_c K)&=&
\frac{G_F}{\sqrt 2}V_{cb}V_{cs}^* \hat a_2\,
\langle M_c |\bar c\gamma^\mu(1- \gamma_5) c|0\rangle
\langle K|\bar s\gamma_\mu(1-\gamma_5) b|B\rangle\,,
\end{eqnarray}
where $G_F$ is the Fermi constant, $V_{ij}$ are the CKM matrix elements,
 $M_c$ represents $J/\psi$ of $J^{PC}=1^{--}$ or 
the exotic $X_c^0$ state  with its constituent being 
$c\bar c q\bar q$.
For simplicity, we take that the quantum numbers of $X_c^0$ are  
$J^{PC}=1^{++}$, such as the established $X(3872)^0$ state.
Note that 
$Y(4140)$, observed in the resonant 
$B^-\to Y(4140)K^-,Y(4140)\to J/\psi \phi$ decay~\cite{Aaltonen:2009tz,Aaltonen:2011at},
is also assumed to be the $J^{PC}=1^{++}$ state and 
 treated as one of the $X_c^0$ states with the tetraquark of $c\bar c s\bar s$~\cite{Stancu:2009ka}.
To calculate the
processes in Fig.~\ref{fig1}, we need to know the matrix elements of 
$\langle X_c^0|\bar c\gamma_\mu(1-\gamma_5) c|0\rangle$, which is the most difficult part unless
these can be related to the observed quantities.
In Eq.~(\ref{eq1}), the parameters $a_2$  and $\hat a_2$, 
 involving  the non-factorizable effects,
can be extracted from the observed branching ratios of 
${\cal B}(\Lambda_b\to J/\psi \Lambda)$ and ${\cal B}(B^-\to J/\psi K^-)$, respectively.
The matrix elements of the $\Lambda_b\to \Lambda$ and $B\to K$
transitions in Eq. (\ref{eq1}) are in the forms of
\begin{eqnarray}\label{twoMX}
\langle \Lambda|\bar s \gamma_\mu b|\Lambda_b\rangle&=&
\bar u_\Lambda \bigg[f_1\gamma_\mu+\frac{f_2}{m_{\Lambda_b}}i\sigma_{\mu\nu}q^\nu+
\frac{f_3}{m_{\Lambda_b}}q_\mu\bigg] u_{\Lambda_b}\,,\;\nonumber\\
\langle \Lambda|\bar s \gamma_\mu\gamma_5 b|\Lambda_b\rangle&=&
\bar u_\Lambda \bigg[g_1\gamma_\mu+\frac{g_2}{m_{\Lambda_b}}i\sigma_{\mu\nu}q^\nu+
\frac{g_3}{m_{\Lambda_b}}q_\mu\bigg]\gamma_5 u_{\Lambda_b}\,,\;\nonumber\\
\langle K|\bar s\gamma_\mu(1-\gamma_5) b|B\rangle&=&
\bigg[(p_B+p_K)^\mu-\frac{m^2_B-m^2_K}{t}q^\mu\bigg]F_1^{BK}(t)+
\frac{m^2_B-m^2_K}{t}q^\mu F_0^{BK}(t)\,,
\end{eqnarray}
with $t\equiv q^2$, where the momentum dependences of
the form factors 
are given by~\cite{Hsiao:2014mua} 
\begin{eqnarray}\label{f1g1}
f_1(t)=\frac{f_1(0)}{(1-t/m_{\Lambda_b}^2)^2}\,,\;
g_1(t)=\frac{g_1(0)}{(1-t/m_{\Lambda_b}^2)^2}\,,
\end{eqnarray}
and~\cite{MS}
\begin{eqnarray}\label{form2}
F^{BK}_1(t)=\frac{F^{BK}_1(0)}{(1-\frac{t}{M_V^2})(1-\frac{\sigma_{11} t}{M_V^2}+\frac{\sigma_{12} t^2}{M_V^4})}\,,\;
F^{BK}_0(t)=\frac{F^{BK}_0(0)}{1-\frac{\sigma_{01}t}{M_V^2}+\frac{\sigma_{02} t^2}{M_V^4}}\;.
\end{eqnarray}
Note that
 the other form factors $f_{2,3}(g_{2,3})$ in Eq.~(\ref{twoMX}) 
that need the loop calculations to flip the valence quark spins have been calculated
to be small and safely ignored.
In terms of the $SU(3)$ flavor and $SU(2)$ spin symmetries, 
one can relate $f_1(0)$ and $g_1(0)$  in Eq.~(\ref{f1g1}) to be~\cite{Hsiao:2015cda}
\begin{eqnarray}\label{Cp}
f_1(0)=g_1(0)=-\sqrt{2/3}\,C_F,
 \end{eqnarray}
with $C_F$ to be extracted from 
the measured $\Lambda_b\to p(K^-,\pi^-)$ decays~\cite{Hsiao:2014mua}.
With $X_c^0$ being $J^{PC}=1^{++}$, the matrix elements in Eq.~(\ref{eq1}) of 
the $0\to J/\psi$ and $0\to X_c^0$ productions can be parameterized as
\begin{eqnarray}\label{fXc}
\langle J/\psi|\bar c\gamma_\mu c|0\rangle&=&m_{J/\psi}f_{J/\psi}\varepsilon_\mu^*\,,\nonumber\\
\langle X_c^0|\bar c\gamma_\mu\gamma_5 c|0\rangle&=&m_{X_c^0}f_{X_c^0}\varepsilon_\mu^*\,,
\end{eqnarray}
where $m_{J/\psi(X_c^0)}$, $f_{J/\psi(X_c^0)}$ and  $\varepsilon_\mu^*$ are the mass,  decay constant 
and polarization for $J/\psi(X_c^0)$, respectively. 
Because of the exotic nature of the $X_c^0$ state, which could be
the $D\bar D^*$ molecule, the hybrid $c\bar c g$ state, or the tetraquark state,
no present QCD model can derive $f_{X_c^0}$. Nonetheless, as we propose that 
$\Lambda_b\to X_c^0 \Lambda$ and $B\to X_c^0 K$ are connected,
we are able to eliminate the unknown $f_{X_c^0}$ and predict 
${\cal B}(\Lambda_b\to X_c^0 \Lambda,X_c^0\to J/\psi \pi^+ \pi^-)$
in terms of the observed ${\cal B}(B\to X_c^0 K, X_c^0\to J/\psi \pi^+ \pi^-)$.

\section{Numerical analysis and Discussions}
For the numerical analysis, the theoretical inputs of
the CKM matrix parameters in terms of 
the Wolfenstein  parameterization are taken to be
$(\lambda,\,A,\,\rho,\,\eta)=(0.225,\,0.814,\,0.120\pm 0.022,\,0.362\pm 0.013)$~\cite{pdg}. 
For the form factor in Eq.~(\ref{Cp}), 
we choose $C_F=0.136\pm 0.009$~\cite{Hsiao:2014mua}, 
which 
is consistent with other QCD model calculations
and used to explain the data in the $\Lambda_b$ decays~\cite{Hsiao:2015cda,Hsiao:2014mua}.
In addition, from Ref.~\cite{MS} we get
$F^{BK}_1(0)=F^{BK}_0(0)=0.36$ with $\sigma_{11}=0.43$, $\sigma_{12}=0$,
$\sigma_{01}=0.70$, $\sigma_{02}=0.27$ and $M_V=5.42$ GeV.
For the parameters $a_2$ ($\hat a_2$), 
we take $(a_2,\hat a_2)=(0.154\pm 0.024,0.268\pm 0.004)$,
which are extracted from 
$\Lambda_b\to J/\psi \Lambda$~\cite{Hsiao:2015cda}
and $B^-\to J/\psi K^-$~\cite{pdg}, respectively.
In terms of Eq.~(\ref{eq1}), we obtain
\begin{eqnarray}\label{R_br}
{\cal R}_{X_c^0}\equiv\frac{{\cal B}(\Lambda_b\to X_c^0 \Lambda)}{{\cal B}(B^-\to X_c^0 K^-)}
&=&0.61\pm 0.20\,,
\end{eqnarray}
where the unkown decay constant $f_{X_c^0}$ has been eliminated.
The measurements for $B^-\to X(3872)^0 K^-$ and $B^-\to Y(4140)^0 K^-$
give~\cite{pdg}
\begin{eqnarray}\label{X}
{\cal B}(B^-\to K^-(X(3872)^0\to)J/\psi \pi^+\pi^-) &=&
(8.6\pm 0.8)\times 10^{-6}
\end{eqnarray}
and~\cite{Aaltonen:2009tz,Aaltonen:2011at}
\begin{eqnarray}\label{Y}
{\cal B}(B^-\to K^-(Y(4140)\to)J/\psi\phi) &=&
(0.149\pm 0.039\pm 0.024){\cal B}(B^-\to J/\psi \phi K^-)
\nonumber\\
&=&(7.7\pm 3.5)\times 10^{-6}
\end{eqnarray}
where we have used
${\cal B}(B^-\to J/\psi \phi K^-)=(5.2\pm 1.7)\times 10^{-5}$~\cite{pdg}.
By relating Eq.~(\ref{R_br}) to Eqs.~(\ref{X}) and (\ref{Y}),
we find
\begin{eqnarray}\label{pre}
{\cal B}(\Lambda_b\to \Lambda(X(3872)^0\to) J/\psi \pi^+\pi^-)=(5.2\pm 1.8)\times 10^{-6}\,,\\
{\cal B}(\Lambda_b\to \Lambda(Y(4140)\to) J/\psi \phi)=(4.7\pm 2.6)\times 10^{-6}\,,
\end{eqnarray}
respectively,
which can be reliable predictions to be compared with the future data.
We remark  that 
${\cal B}(\bar B^0\to \bar K^0(X(3872)^0\to)J/\psi \pi^+\pi^-)=(4.3\pm 1.3)\times 10^{-6}$~\cite{pdg}
can also lead to  similar results but with larger uncertainties than those in Eq.~(\ref{pre}).
%
It should be noted that 
 the quantum numbers for $Y(4140)$ have  not been  experimentally identified yet, although
they are predicted to be 
$J^{PC}=$ $0^{++}$ ($2^{++}$) in Ref.~\cite{Liu:2009ei} and $1^{-+}$ in Ref.~\cite{Mahajan:2009pj}
besides $1^{++}$ in Ref.~\cite{Stancu:2009ka}.   
We emphasize  that, 
even it is finally measured to have $J^{PC}=0^{++}$~\cite{Wang:2014gwa} or $1^{-+}$,  
the decay of $\Lambda_b\to \Lambda(Y(4140)\to) J/\psi \phi$ 
can still be examined by our method.
However, 
the factorization approach would not support
the tensor (T) identification of the $J=2$ state due to 
$\langle T|\bar c\gamma_\mu(1-\gamma_5) c|0\rangle=0$.

Finally, we note that unlike  
$B^-\to X(3872)^0 K^-$,
which receives the dominant contribution from the doubly charmful $b\to c\bar c s$ transition,
the decay of $B^-\to X(3872)^- \bar K^0$ is forbidden in Fig.~\ref{fig1} as supported by the experiment due to 
its non-observation~\cite{Aubert:2004zr}, where $X(3872)^-$ is the charged counterpart of $X(3872)^0$.
However, this mode can proceed from the charmless $b\to d\bar d s$ transition,
provided that the $c\bar c$ contents in $X(3872)^-$ come from the intrinsic charm within the $B$ meson,
which is similar to the pentaquark state productions 
in the $\Lambda_b$ decays~\cite{Hsiao:2015nna,Aaij:2015tga}.
As a result, in the charmless $B$ decays, 
the branching ratios of the three possible exotic decays of
$\bar B^0\to X_c^+ K^-,X_c^+ \pi^-$, and $B^-\to X_c^- \bar K^0$ can be at the same level.
In addition, 
the intrinsic charm mechanism would be used to the productions of the charged 
Y and Z particles as $\bar B^0\to Z(4430)^+K^-$ with 
$Z(4430)^+$ to consist of $c\bar c u \bar d$~\cite{Aaij:2014jqa,Brodsky:2015wza}.
Moreover, the analogous statements for the corresponding $\Lambda_b$ decays can also be drawn.

\section{Conclusions}
We have explored the possibility to find the exotic meson states, such as  
the tetraquark four-quark bound states of $X_c^0=c\bar c u\bar u(d\bar d)$ and $c\bar c s\bar s$
in the $\Lambda_b$ decays.
In particular, by concentrating on the scenarios with $X(3872)^0$ and $Y(4140)$
being $J^{PC}=1^{++}$, we have studied 
the doubly charmful $\Lambda_b\to X_c^0\Lambda$  decays.
By connecting $\Lambda_b\to \Lambda X_c^0$ to $B^-\to K^- X_c^0$,
we have found that
${\cal B}(\Lambda_b\to \Lambda(X(3872)^0\to) J/\psi \pi^+\pi^-)$ and ${\cal B}(\Lambda_b\to \Lambda(Y(4140)\to) J/\psi \phi)$
are $(5.2\pm 1.8)\times 10^{-6}$ and $(4.7\pm 2.6)\times 10^{-6}$, respectively.
As these predicted branching ratios  are accessible to the experiments
at the LHCb, a measurement  will be the first clean experimental evidence for the $XYZ$ states in baryonic decays.

\section*{ACKNOWLEDGMENTS}
The work was supported in part by 
National Center for Theoretical Sciences, 
National Science Council (NSC-101-2112-M-007-006-MY3), 
MoST (MoST-104-2112-M-007-003-MY3) and 
National Tsing Hua University (104N2724E1).


\begin{thebibliography}{99}
\bibitem{QN_LHCb} 
R.~Aaij {\it et al.} [LHCb Collaboration],
Phys.\ Rev.\ Lett.\  {\bf 110}, 222001 (2013); Phys.\ Rev.\ D {\bf 92}, 011102 (2015).

\bibitem{XYZ_Rev}
H.~X.~Chen, W.~Chen, X.~Liu and S.~L.~Zhu,
  arXiv:1601.02092 [hep-ph].

\bibitem{Maiani:2004vq} 
L.~Maiani {\cal it al.}, 
Phys.\ Rev.\ D {\bf 71}, 014028 (2005).

\bibitem{Stancu:2009ka} 
F.~Stancu, 
J.\ Phys.\ G {\bf 37}, 075017 (2010).

\bibitem{pdg}
K.A.~Olive {\it et al.}  [Particle Data Group Collaboration], Chin.\ Phys.\ C {\bf 38}, 090001 (2014).

\bibitem{Bala:2015wep} 
A.~Bala {\it et al.} [Belle Collaboration],
Phys.\ Rev.\ D {\bf 91}, 051101 (2015).

\bibitem{molecule_A} 
E.~Braaten and M.~Kusunoki, Phys.\ Rev.\ D {\bf 69}, 074005 (2004);
E.~S.~Swanson, Phys.\ Lett.\ B {\bf 588}, 189 (2004);
N.A.~Tornqvist, Phys.\ Lett.\ B {\bf 590}, 209 (2004);
M.B.~Voloshin, Phys.\ Lett.\ B {\bf 604}, 69 (2004).

\bibitem{Li:2004sta} 
B.A.~Li, Phys.\ Lett.\ B {\bf 605}, 306 (2005).

\bibitem{Hsiao:2015cda} 
Y.K.~Hsiao, P.Y.~Lin, C.C.~Lih and C.Q.~Geng,
Phys.\ Rev.\ D {\bf 92}, 114013 (2015).

\bibitem{FA_Xc}
C.M.~Zanetti, M.~Nielsen and R.D.~Matheus,
Phys.\ Lett.\ B {\bf 702}, 359 (2011).

\bibitem{Aaltonen:2009tz} 
T.~Aaltonen {\it et al.} [CDF Collaboration], Phys.\ Rev.\ Lett.\  {\bf 102}, 242002 (2009).

\bibitem{Aaltonen:2011at} 
T.~Aaltonen {\it et al.} [CDF Collaboration], arXiv:1101.6058 [hep-ex].
 



\bibitem{Hsiao:2014mua} 
Y.K.~Hsiao and C.Q.~Geng, Phys.\ Rev.\ D {\bf 91}, 116007 (2015).

\bibitem{MS} D. Melikhov and B. Stech, Phys. Rev. {\bf D62}, 014006 (2000).

\bibitem{Liu:2009ei} 
X.~Liu and S.L.~Zhu,
Phys.\ Rev.\ D {\bf 80}, 017502 (2009) [Phys.\ Rev.\ D {\bf 85}, 019902 (2012)].

\bibitem{Mahajan:2009pj} 
N.~Mahajan, 
Phys.\ Lett.\ B {\bf 679}, 228 (2009).

\bibitem{Wang:2014gwa} 
Z.G.~Wang, Eur.\ Phys.\ J.\ C {\bf 74}, 2963 (2014);
arXiv:1601.05541 [hep-ph].



\bibitem{Aubert:2004zr} 
B.~Aubert {\it et al.} [BaBar Collaboration], Phys.\ Rev.\ D {\bf 71}, 031501 (2005).


\bibitem{Hsiao:2015nna} 
Y.K.~Hsiao and C.Q.~Geng, Phys.\ Lett.\ B {\bf 751}, 572 (2015).

\bibitem{Aaij:2015tga} 
R.~Aaij {\it et al.} [LHCb Collaboration], Phys.\ Rev.\ Lett.\  {\bf 115}, 072001 (2015).
 
\bibitem{Aaij:2014jqa} 
R.~Aaij {\it et al.} [LHCb Collaboration],
Phys.\ Rev.\ Lett.\  {\bf 112}, 222002 (2014).

\bibitem{Brodsky:2015wza} 
S.J.~Brodsky and R.F.~Lebed,
Phys.\ Rev.\ D {\bf 91}, 114025 (2015).

\end{thebibliography}
\end{document}